\documentclass[conference]{IEEEtran}
 \usepackage{graphicx}
 \usepackage{fixltx2e}
 \usepackage{longtable}
 \usepackage{tabularx}
\usepackage{epsfig}
\usepackage{subfigure}
\usepackage{multirow}
\usepackage{bigstrut}

\ifCLASSINFOpdf
\else
\fi
\begin{document}

%
\title{Efficient Cooperative Anycasting for AMI Mesh Networks}

%
%
%

\author{Sedat Gormus, Mohammud Zubeir Bocus \\
        Telecommunications Research Laboratory, Toshiba Research Europe Ltd.,\\
Bristol, BS1 4ND, UK\\
Email:\{sedat.gormus, zubeir.bocus\}@toshiba-trel.com}

\maketitle
\thispagestyle{empty}
\pagestyle{empty}

\begin{abstract}
We have, in recent years, witnessed an increased interest towards enabling a Smart
Grid which will be a corner stone to build sustainable energy efficient communities. 
An integral part of the  future Smart Grid will be the communications infrastructure
which will make real time control of the grid components possible. Automated Metering
Infrastructure (AMI) is thought to be a key enabler for monitoring and controlling the customer loads. 
This paper proposes an efficient cooperative anycasting approach for wireless mesh networks with the aim of achieving reduced traffic and increased utilisation of the network resources. The proposed cooperative anycasting has been realised as an enhancement on top of the Routing Protocol for Low Power and Lossy
Networks (RPL), a connectivity enabling mechanism in wireless AMI mesh networks. In this protocol, smart meter nodes utilise  an anycasting approach to facilitate efficient transport of metering data to the
concentrator node. Moreover, it takes advantage of a distributed approach ensuring scalability.
\end{abstract}


%
\IEEEpeerreviewmaketitle

\section{Introduction}
%
%
%
%

\IEEEPARstart{M}{otivated} by the need to improve energy efficiency
and reduce the environmental impact, we observe a push towards
enabling a `Smart Grid' recently. One of the corner stones of the Smart Grid
is the Automated Metering Infrastructure (AMI), planned to be a
network consisting of meter and concentrator nodes. In this set-up, meter nodes at the consumer sites transport their readings automatically to the Meter Data Management System (MDMS) through concentrators
installed by the access network provider.

A broad range of communication technologies can be considered to realise a metering infrastructure. It is likely to be a wide mix of public and private, wired and wireless, standard and proprietary networking solutions. In a neighborhood area network (NAN), WiFi, GPRS and WiMAX are some
of the example technologies being considered for this segment. On the other hand, in home area networks (HAN), IEEE 802.15.4 (Zigbee) is the main wireless technology that is being considered, followed by WiFi and power line communications (PLC). At the moment there is no clear winner technology yet. It is possible that we will see a mixture of different technologies in order to realise smart metering networks. In  this paper, we focus on a wireless mesh networking solution because of its ease of deployment, reliability features, and cost effectiveness.

The topology in such a network typically takes the form of a number of trees, each terminated at a concentrator node. The meter nodes transmit  meter readings through multi-hop radio links to a concentrator in close proximity. In an urban deployment, there could be a large number of flats/houses with one or more meters (e.g., electricity, water, gas) in each of these. In such a scenario, the metering communication may sufeer from interference cause the neighbouring nodes. We have shown in our previous work \cite{orpl} that it is possible to improve the scalability and the utilisation of the network through 
efficient protocol design that make use of anycasting \cite{anycast}. In this case, with lossy
wireless links, the probability of receiving a packet at any node in a neighbour-set increases with increasing number of nodes in this set. Furthermore, we have shown that an optimal anycast set can be straightforwardly found in a network utilising RPL routing protocol \cite{draft-ietf-rpl}, since the neighbour set size in such a network has to be small due to the memory limitation on the sensor devices. RPL is a route selection and maintenance  mechanism designed for low power and lossy networks (LLNs), standardised by the IETF ROLL working group, and is a good candidate for enabling connectivity in AMI mesh networks. 

In \cite{orpl} and \cite{orpl-tpds}, we have shown that an adaptive cooperative communication mechanism that takes advantage of cross layer design can reduce network retransmissions by 2 to 3 fold as compared to the standard RPL implementation. On the other hand, we also highlighted that such a cooperative communication protocol (called ORPLx) suffers in terms of packet delivery  when the network load is high. This is mainly due to interference caused within the network as a result of concurrent transmissions of non-neighbouring nodes (i.e, hidden terminal problem). In this paper, we address this problem by introducing interference estimation into the cooperative anycasting protocol. Furthermore, we implement two link interference estimation methods involving active probing and passive monitoring and analyse their performance and protocol overhead in different network scenarios.

In this paper, we outline the details of a low complexity adaptive cooperative anycasting mechanism
for RPL routing protocol by modifying MAC and Network layers of the protocol stack. In section \ref{protocolOutline}, the opportunistic communication approach for RPL is briefly summarised along with the analytical explanation of the adaptive cooperative anycasting extensions. In section \ref{protocolImplementation}, we outline the implementation details of the adaptive ORPLx protocol for the Contiki operating system \cite{contiki}. The protocol performance is analysed in Section \ref{evaluation}. Section \ref{conclusion} highlights the conclusions drawn from this study.

\section{Opportunistic Communication for RPL Routing}\label{protocolOutline}


The proposal described in the RPL standard \cite{draft-ietf-rpl}
creates a Directed Acyclic Graph (DAG) rooted at the sink
(concentrator in this case) to maintain network state information. A
path from a smart meter node orienting towards and terminating at
the concentrator (sink node) consists of the edges in the DAG. Each
node in the DAG is associated with a rank value such that the rank
of nodes along any path to the DAG root should be monotonically
decreasing. In order to construct a DAG, the root node will issue a
control message beacon (called DAG Information Object (DIO)).
Any node that receives a DIO message and is willing to join the DAG
should add the DIO sender (the previous node traversed by the DIO)
to its parent list, compute its own rank (associated with the parent
node) and broadcast the DIO message with the updated rank
information.

ORPL and ORPLx introduced in \cite{orpl-tpds} makes use of 
the existing parent structure of the RPL protocol which stores 
one or more candidate parents (if possible) at each 
client node in addition to the default parent. 
RPL forwards the frame to the default parent alone.
Although other parents within radio range may overhear the transmission, they will simply ignore it.
ORPL takes advantage of this feature. The client node forwards the frame both to the default parent and the candidate parent. If the default parent
receives the frame, it forwards the frame to its parents and
the candidate parents discard the copy of the forwarded frame when they detect the default parent transmission. On the other hand, if the default parent fails to receive the frame, the highest priority candidate parent acknowledges the frame and forwards it to the next hops.

ORPL uses overhearing-based coordination to minimise the control
overhead. The source node selects its parents such that the parent nodes 
are within the wireless range of each other. Furthermore, in order to increase the efficiency of the protocol, only the highest priority parent, which receives the client's frame successfully, sends an acknowledgement frame to the source node. The other nodes in the parent set which have received the frame successfully discards the buffered frame when they overhear the acknowledgement frame. This approach has been proven to be more efficient in DAG based sensor networks \cite{orpl-tpds} as compared to ExOR \cite{Biswas2004} like protocols where each and every parent nodes has to acknowledge the successfully received frame.    

\begin{figure}
 \centering
   \includegraphics[width=0.45\textwidth]{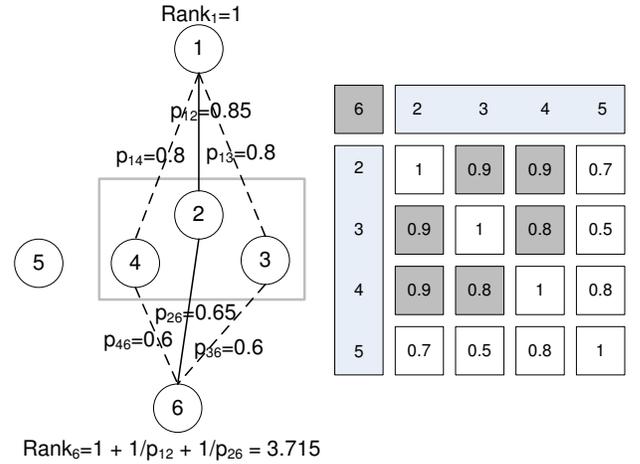} 
    \caption{A simple RPL topology and corresponding neighbourhood map
    for node 6. The rank value of node 6 corresponds to the minimum
    shortest path route (6-2, 2-1).}
    \label{fig.topologymap}
\end{figure}

\begin{figure}
 \centering
   \includegraphics[width=0.4\textwidth]{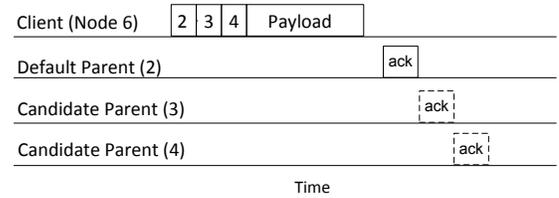}
\caption{Timeline for the forwarding process. The frame is acknowledged and
forwarded by the default parent. The candidate parents forward their
frames when they can not detect the ack frame from the parent with higher priority.}
\label{fig.timeline}
\end{figure}

For example, node 6 in Fig.\ref{fig.topologymap} selects a parent set consisting of nodes 2,3 and 4 since they provide the highest reduction in terms of end-to-end communication cost. The default parent node (node 2) acknowledges and forwards the frame upon a successful reception. If node 2 fails to receive the frame transmitted by node 6, the highest priority candidate parent (node 3 in this case) forwards the frame to the sink node as highlighted in Fig.\ref{fig.timeline}.   

While ORPL can significantly reduce the number of retransmissions in an RPL network \cite{orpl}, its overhead due to cooperation errors increases in the scenarios where the communication links suffer from high loss rates. In this case, the retransmission overheads are mainly caused by the dropped acknowledgement frames from the forwarding parent to the client node. To minimize this overhead, we have come up with ORPLx \cite{orpl-tpds} protocol where the MAC retransmission limits were calculated using the anycast success probability of the links from the client node $i$ to the parent set $S(i)$. For each node $j\in S(i)$, $p_{ij}$ is the probability of correctly receiving the message from the source node without any interference. The anycast link probability can be expressed as;
\begin{equation}
    p_A = 1-\prod_{j \epsilon S(i)}(1-p_{ij}). \label{eq:pA}
\end{equation}

Here, we can introduce a target anycast link success rate and calculate the MAC retransmission limit to achieve the desired level of link reliability. The MAC layer retry limit for a given link success probability $p_t$ is given as \cite{orpl-tpds}; 

\begin{equation}
\sum_{n=1}^{k}{{k\choose q} p_A^n (1-p_A)^{k-n}} = 1- (1-p_A)^k \geq
p_t, \label{eq:1}
\end{equation}

where $p_A$ is the anycast probability and $k$ is the number of transmission to achieve a link success rate greater than $p_t$. By taking the logarithm at both sides
of the inequality we obtain the minimum number of retransmissions
required to meet the target success probability as a function of the
anycast probability:

\begin{equation}
k_{\textrm{min}}=\max\left\{1,\left\lceil\frac{\log(1-p_t)}{\log(1-p_A)}\right\rceil\right\}.
\label{eq:kmin}
\end{equation}

While this approach eliminates most of the duplicate retransmissions due to ACK frame failures, the accurate estimation of anycast link success rate poses a real challenge since packet drops not only happen due to channel errors, but also due to collisions in the network. Furthermore, when a collision happens, the entire parent set of the transmitting client node fails to receive the transmitted frame. Hence, in a high contention network ORPLx protocol struggles to achieve the target PDR when the anycast link success probabilities are based on the channel loss statistics. In this case, the formula given in Eq. (\ref{eq:kmin}) does not provide an accurate MAC retry limit. To overcome this problem, we considered two approaches. The first approach uses probing frames to monitor and estimate the anycast link statistics. The second approach takes a passive listening approach where the anycast link RSSI values are mapped to link success rates and the collisions reported by the 802.15.4 PHY is factored into the formula given ins Eq. \ref{eq:kmin}.

\section{Adaptive anycasting extensions for ORPL Protocol}

As mentioned earlier, the minimum number of MAC retries in a cooperative anycasting scenario depends on several factors such as the number of nodes in the parent set, the individual link success probabilities from the source node to parent nodes and the collisions in the network due to contention between the nodes in the network using a Carrier Sense Multiple Access (CSMA) type MAC layer. In this case, individual link success probabilities should be estimated correctly in order to calculate $k_{min}$ value accurately. 

The straight forward approach to gathering link statistics is to use probe frames, but this method introduces extra control overheads into the sensor network and may use significant amount of bandwidth resources. On the other hand, if we can make use of freely available RSSI and collision statistics to calculate $k_{min}$ value, this may potentially provide a better solution compared to probing. This approach requires reformulating the anycast link probability since RSSI does not provide reliable loss statistics in the presence of collisions. 

We want to calculate the minimum number of retransmissions
required to guarantee that at least one node in the set receives the
message correctly with probability not smaller than a target $p_t$ in the presence of collisions with a collision probability $p_c$. The anycast probability, i.e., the probability of
at least one node receiving the message correctly after a single
transmission is given in Eq. (\ref{eq:pA}). Let us define the successful reception probability of a packet for perfect channel condition with collision probability $p_c$ as $\hat{p_c}$;
\begin{equation}
    \hat{p_c} = 1-p_c. \label{eq:pC}
\end{equation}

Given these equations, a frame can only be successfully transmitted when there is no collision in the network and anycast transmission of the frame is successful with the assumption that the packet drops due to interference and the channel errors are statistically independent. In this case, the link success probability of the link depicted by $p_{Ac}$ becomes;

\begin{equation}
    p_{Ac} = p_A * \hat{p_c}. \label{eq:pAc}
\end{equation}
 
By substituting (\ref{eq:pAc}) into (\ref{eq:kmin}), we can easily find the minimum number of transmissions required to achieve a link success rate of $p_t$ in the presence of channel errors and collisions.

\section{Analysis of the protocol using Contiki OS}\label{protocolImplementation}
We used TelosB \cite{telosb} devices with open source Contiki operating system as the development platform to implement MAC layer adaptive anycasting with ORPL protocol whose implementation details are given in \cite{orpl}. The performance of the proposed adaptive MAC layer modifications is analysed through the Cooja emulator \cite{contiki}  which facilitates development, testing and debugging of the code before running it on the target platform.

\subsubsection{MAC Layer Modifications}
The link quality between the client node and its parents can be estimated by using an empirical approach where the RSSI value of the received signal can be mapped to PDR levels. Here, it is assumed that there is no interference or collisions in the system. To achieve this, each client node in a sensor network stores the average received power levels for all the nodes in its parent set where an exponential averaging process\footnote{$RSSI_{avg}=\alpha*RSSI_{current} + (1-\alpha)*RSSI_{avg}, \alpha = 0.25$} can be used. As we highlighted earlier, using only RSSI for anycast link quality estimation is reliable in the scenarios where there is no or limited collisions in the network. On the other hand, we can straightforwardly introduce collisions into the MAC layer retransmission limit calculation by modifying the anycast probability formula. However, the logarithm function in Eq. (\ref{eq:kmin}) is not easy to implement on a sensor hardware. In such a case, Eq. (\ref{eq:5}),  which represents the first two terms of power series expansion of Eq. (\ref{eq:kmin}), can be used for calculating the number of retries for a target link success rate of $p_t$;

\begin{equation}
\label{eq:5}
    k_{\textrm{f}} = \left\{
    \begin{array}{ll}
        1, &\mbox{for } \Theta < 1\\ \\
        \left\lfloor\Theta + 1 \right\rfloor, &\mbox{for } 1 \leq \Theta < 1.5\\ \\
        \left\lceil\Theta + 1 \right\rceil, &\mbox{otherwise} \\
    \end{array} \right.,
\end{equation}
where
\begin{equation}
\Theta = \left[\frac{p_t+\frac{p_t^2}{2}}{p_{Ac}+\frac{p_{Ac}^2}{2}}\right].
\end{equation}

When a client node needs to send a data frame to the concentrator
node, it needs to calculate the retransmit limit for the target link
success threshold $p_t$. Firstly, the client node retrieves stored
received power levels of the entire anycast set. Secondly, the collision statistics reported by the 802.15.4 hardware are retrieved. Thirdly, the empirical link statistics derived from the received power levels are used to calculate the collision free anycast link success probability. Consequently, the client node uses collision statistics to calculate the final link success probability and feeds this value to Eq. (\ref{eq:5}) to calculate $k_f$. The $k_f$ value represents the minimum number of MAC retransmissions to achieve the desired level of link reliability.

As shown in Eq. (\ref{eq:5}), we use a retransmission limit of 1
only when the estimated number of retransmissions is less than 1.
This means that the transmission of the frame will be successful
with a probability greater than the target success probability
$p_t$. On the other hand, when the estimated retransmit limit is
greater than 1, an extra transmission is performed upon an ACK failure.

\section{Performance Results}\label{evaluation}
\subsection {Emulation Scenario}
The performance of adaptive anycasting protocol extensions is evaluated against RPL, ORPL and ORPLx protocols using a 20 node planned network deployment and an example smart metering deployment with map data belonging to a neighbourhood in Bristol, UK. The scenarios are;
\begin{itemize}
\item{Scenario 1: \emph{Planned deployment of 20 nodes}}
The objective of this scenario is to analyse the performance of the new protocol extensions as compared to ORPLx, ORPL and RPL protocols in terms of PDR and MAC layer retransmissions. In this scenario, each client node can select a maximum of three anycast parents. Each client node periodically transmits a meter reading frame of 60 bytes to the concentrator. The scheduling of the readings for this scenario is done by dividing the meter reading interval into 20 non overlapping time slots. When a client node fails to transmit a frame due to an ACK failure, the MAC layer schedules the frame with a linear back
off timer as per CSMA MAC requirements.

\item{Scenario 2:  \emph{Example deployment with map data}}
This scenario comprises of a planned deployment of 120 client nodes that utilises the map data of a neighborhood located in city of Bristol, UK. It is assumed that each house is equipped with a smart electricity meter. The objective of this scenario is to analyse the performance of the proposed extensions in a realistic deployment scenario.

\end{itemize}

In each scenario, two sets of simulations were carried out for high and low contention scenarios. While the frame drops occur due to interference and link losses in the high contention case, the errors generally occurs due to channel impairments in the low contention scenario. The results of Scenario 1 represent the network performance for different link success probabilities ranging from 70 to 90 percent success rate which depends on the distance as given by the free-space path loss formula \cite{balanis}. For example, a scenario with link success rate of 70\% can be visualised as one in which two nodes are at the maximum distance that enables them to talk to each other.
On the other hand, when the nodes are just besides each other ( less than 1 meter apart), the actual link success rate becomes around 100\% for the same scenario. The results presented show the aggregate values of the performance metrics (considered in this study) for the whole network after 100 frame transmissions. ORPLx target packet delivery ratio is set to 99\%. The probing interval was set to an interval between 20 and 25 seconds for the case where forward link quality is estimated using probe packets. Each client node sends a 16 bytes ICMPv6 probe packet and the exact time of the probe packet depends on the processor load of the client node. For example, when the client node has no scheduled job, it sends a probe packet to each of its parents within the probing interval. Furthermore, when ORPL is used, the number of retransmissions represents the transmissions due to packet failures plus the retransmissions due to cooperation errors.

\subsection{Performance Results}


\begin{figure}
 \centering
   \includegraphics[width=0.39\textwidth]{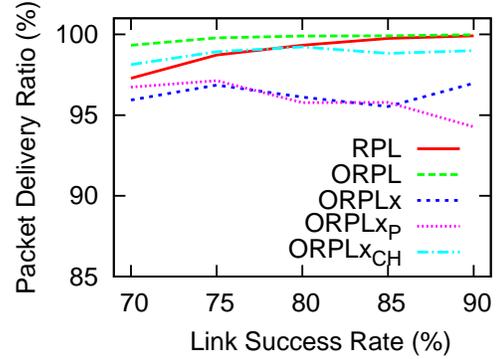}%
\caption{High contention PDR in scenario 1}
\label{fig.pdr20_5}
\end{figure}

\begin{figure}
 \centering
  \includegraphics[width=0.39\textwidth]{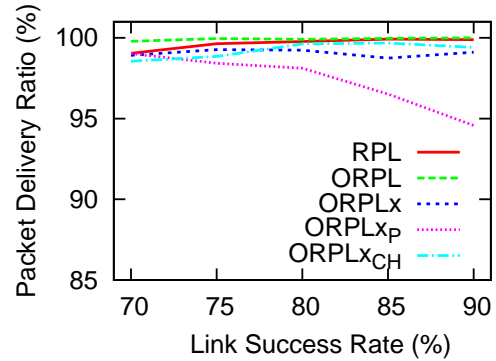}
\caption{Low contention PDR in scenario 1}
\label{fig.pdr20_15}
\end{figure}

The PDR results in Fig. \ref{fig.pdr20_5} and \ref{fig.pdr20_15} indicate that using probing (ORPLx\textsubscript{P}) for forward channel estimation does not provide any significant performance improvement in both 5 seconds high contention and 15 seconds low contention scenarios. While probing can reliably estimate the link loss probability when the network is not saturated (15 seconds), it fails to capture collisions due to infrequent probing intervals. Furthermore, for the high contention scenario with 5 seconds frame interval, probing introduces extra traffic to the network and fails to show any performance benefit. On the other hand, making use of collision statistics available at the 802.15.4 hardware can provide a significant performance gain in high contention scenarios where the main source of packet drops is the interference. It can be seen from the results in Fig. \ref{fig.pdr20_5} and \ref{fig.pdr20_15} that ORPLx taking advantage of collision history (ORPLx\textsubscript{CH}) achieves an overall PDR performance very close to the target PDR of 99\%.

\begin{figure}
 \centering
   \includegraphics[width=0.39\textwidth]{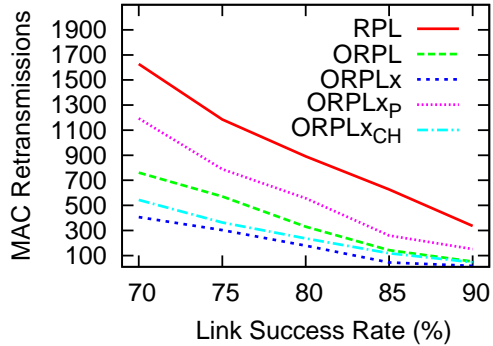}
\caption{Number of retransmissions for 5 seconds frame interval}
\label{fig.rexmit5}
\end{figure}

\begin{figure}
 \centering
   \includegraphics[width=0.39\textwidth]{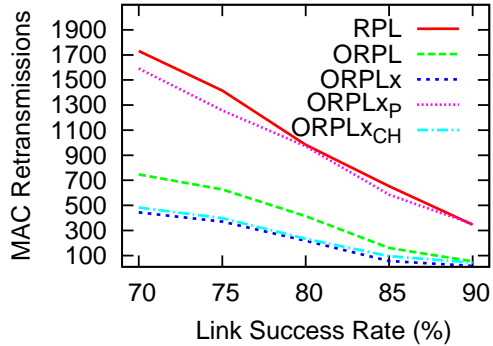}
\caption{Number of retransmissions for 15 seconds frame interval}
\label{fig.rexmit15}
\end{figure}

The MAC layer retransmissions are given in Fig. \ref{fig.rexmit5} and \ref{fig.rexmit15} for high and low contention scenarios. The results show that probing introduces a significant overhead to the network and increases the number of retransmissions as compared to ORPL and ORPLx. The cumulative number of the retries depends on the meter reading interval since all the simulations are run for 100 frame transmissions. In this case, there are less number of probe frames transmitted in the high contention scenario due to frequency (20-25 seconds) of the probe frames. 

On the other hand, ORPLx\textsubscript{CH} has a slightly higher retransmission count as compared to ORPLx in both scenarios. And the retransmission count of ORPLx\textsubscript{CH} is higher in the high contention scenarios compared to low contention scenarios.  ORPLx\textsubscript{CH} takes collisions into account and adjusts the retry limits at the MAC layer. This translates to a better PDR in high contention conditions as compared to ORPLx at the expense of some increase in network traffic. If we look at Fig. \ref{fig.rexmit5}, we see that ORPLx\textsubscript{CH}  requires on average 20-30\% more MAC retransmissions as compared to ORPLx. Whereas for the low contention scenarios, its retransmission performance is very similar to ORPLx which confirms our observation on packet drops due to interference.

\begin{table}[htb]
\caption{PDR and MAC Retransmissions Results for the Example Deployment Scenario }
\label{tab.exp-deployment}
\begin{center}
\begin{tabular}{|c|c|c|c|c|}
\hline &RPL & ORPL & ORPLx & ORPLx\textsubscript{CH} \\
\hline PDR(\%)-15s & 94.34 & 98.07 & 94.04 & 98.5  \\
\hline MAC Tx-15s & 9230 & 6220 & 5217 & 5397 \\
\hline PDR(\%)-30s & 98.89 & 99.25 & 96.73 & 99.7  \\
\hline MAC Tx-30s & 7592 & 5446 & 4159 & 4238 \\
\hline
\end{tabular}
\end{center}
\end{table}

PDR and MAC layer retransmission results for Scenario 2 in Table \ref{tab.exp-deployment} show a similar trend to that of the 20-node scenario. While ORPLx fails to sustain the desired PDR level of 99\% for high contention scenarios (15 seconds metering interval),  ORPLx\textsubscript{CH} achieves the highest PDR result in this scenario. This can be attributed to fact that  ORPLx\textsubscript{CH} adaptively monitors the collisions and channel statistics and uses minimum number of MAC retries to guarantee the reliable delivery of the metering data. This results in a lower retransmission contention and enables the reliable transfer of the metering data. These results shows that  ORPLx\textsubscript{CH} can potentially enable denser AMI deployments and provide a scalable solution.

In summary, using active probing in low throughput wireless mesh networks to estimate link quality is not viable due to overhead and its inability to reliably detect the packet collisions. On the other hand, passive monitoring using 802.15.4 collision statistics can potentially provide the desired levels of traffic reduction in the wireless mesh network for the target PDR levels when such a cooperative anycasting scheme is used.

\section{Conclusions} \label{conclusion}

Wireless mesh networks are low cost, easy to setup and maintain and therefore attractive for realising a practical AMI deployment. In this paper, we proposed an adaptive cooperative anycasting mechanism to improve the efficiency of data transport in AMI mesh networks. We studied its performance through  a Contiki OS implementation and confirmed its superior performance through emulation based experiments as compared to previously proposed protocols. Results from this study have demonstrated the effectiveness of cooperative anycasting in DAG based networks. We conclude that using cooperative anycasting can be a way forward to improving communication efficiency and reliability in wireless sensor networks where nodes are limited in communication bandwidth and experience high losses.




\bibliographystyle{IEEEtran}
\bibliography{ORPLx-Globecom}


%

%
%
%
%
%

\ifCLASSOPTIONcaptionsoff
  \newpage
\fi

\end{document}